\documentclass[iop,apj,tighten,numberedappendix]{emulateapj}
\usepackage{amssymb}
\usepackage{epsfig}
\usepackage{graphics}
\usepackage{natbib}
 
\newcommand {\msun}{M_{\sun}}
\newcommand {\ergs}{{\rm erg\ \rm s^{-1}}}
\newcommand{\bmath}[1]{\mbox{ \boldmath $\!#1\!$ \unboldmath}}

\begin{document}
% \date{}

%%%%%%%%%%%%%%%%%%%%%%%%%%%%%%%%%%%%%%%%%%%%%%%%%%%%%%%%%%%%%%%%%%%%%%%%%%%%%%
%% Title Details and Page Header                                            %%
%%%%%%%%%%%%%%%%%%%%%%%%%%%%%%%%%%%%%%%%%%%%%%%%%%%%%%%%%%%%%%%%%%%%%%%%%%%%%%
\title{A reflection model for the cyclotron lines in the spectra of X-ray pulsars}

\shorttitle{A reflection model for the cyclotron lines in X-ray pulsars}

\author{
 Juri~Poutanen\altaffilmark{1}, 
 Alexander~A.~Mushtukov\altaffilmark{1,2,3},
Valery~F. Suleimanov\altaffilmark{4,5},
Sergey~S.~Tsygankov\altaffilmark{6,1,7}, 
Dmitrij~I.~Nagirner\altaffilmark{2,1}, 
Victor~Doroshenko\altaffilmark{4}, and
Alexander~A.~Lutovinov\altaffilmark{7} 
}

\shortauthors{Poutanen et al.}

\affil{$^1$ Astronomy Division, Department of Physics, PO Box 3000,
FI-90014 University of Oulu, Finland;  juri.poutanen@oulu.fi \\ 
$^2$ Sobolev Astronomical Institute, Saint Petersburg State University,
  Saint-Petersburg 198504, Russia \\
$^3$ Pulkovo Observatory of Russian Academy of Sciences,
  Saint-Petersburg 196140, Russia \\
$^4$ Institut f\"ur Astronomie und Astrophysik, Kepler Center for Astro and Particle Physics, 
Universit\"at T\"ubingen, Sand 1, 72076 T\"ubingen, Germany \\
$^5$ Kazan (Volga region) Federal University, Kremlevskaja str., 18, Kazan 420008, Russia \\
$^6$ Finnish Centre for Astronomy with ESO (FINCA), University of Turku,  V\"ais\"al\"antie 20,
FI-21500 Piikki\"o, Finland \\
$^7$ Space Research Institute of the Russian Academy of Sciences, Profsoyuznaya Str. 84/32, Moscow
  117997, Russia}

%\maketitle

%%%%%%%%%%%%%%%%%%%%%%%%%%%%%%%%%%%%%%%%%%%%%%%%%%%%%%%%%%%%%%%%%%%%%%%%%%%%%%
%% Abstract, Keywords and contact details                                   %%
%%%%%%%%%%%%%%%%%%%%%%%%%%%%%%%%%%%%%%%%%%%%%%%%%%%%%%%%%%%%%%%%%%%%%%%%%%%%%%

\begin{abstract}
Cyclotron resonance scattering features observed in the spectra of some X-ray pulsars show 
significant changes of the line energy with the pulsar luminosity. 
At high luminosities, these variations are often associated with the onset and growth of the  
accretion column, which is believed to be the origin of the observed emission and of the cyclotron lines. 
However, this scenario inevitably implies large gradient of the magnetic field strength 
within the line-forming region, which makes the formation of the observed line-like features problematic. 
Moreover, the observed variation of the cyclotron line energy is
much smaller than could be anticipated for the corresponding luminosity changes.
We argue here that a more physically realistic situation is that the cyclotron
line forms when the radiation emitted by the accretion column is reflected from
the neutron star surface, where the gradient of the magnetic field strength
is significantly smaller. We develop here the reflection model and apply it 
to explain the observed variations of the   cyclotron line energy  
in a bright X-ray pulsar V~0332+53 over a wide range of luminosities.
\end{abstract}


%%%%%%%%%%%%%%%%%%%%%%%%%%%%%%%%%%%%%%%%%%%%%%%%%%%%%%%%%%%%%%%%%%%%%%%%%%%%%%
%% Introduction                                                             %%
%%%%%%%%%%%%%%%%%%%%%%%%%%%%%%%%%%%%%%%%%%%%%%%%%%%%%%%%%%%%%%%%%%%%%%%%%%%%%%
\section{Introduction}
\label{intro}

X-ray pulsars are neutron stars in binary systems accreting matter usually from a massive companion. 
These neutron stars have a sufficiently strong magnetic field, which channels accreting gas
towards magnetic poles. 
Strong magnetic field modifies the observed X-ray spectrum often
manifesting as the line-like absorption  features, the so-called cyclotron lines. 
Such cyclotron resonance scattering features (CRSF), 
sometimes also with harmonics, are observed 
in the spectra of  several X-ray pulsars \citep{Coburn2002,Fil2005,CW12}. 

In some cases, the luminosity related changes of the line energy are observed,
suggesting that  configuration of the line-forming region depends on the
accretion rate. The line energy has been reported to be positively (in
relatively low-luminosity sources; see \citealt{Staub2007,Kloch2012}) 
and negatively-correlated with luminosity (in high-luminosity sources; see \citealt{Mih1998,TsL2006}), 
as well as uncorrelated with it \citep{CP13}.  

This diversity is yet to be explained, and in this work we will focus only on
the high-luminosity case. The negative correlation of the CRSF energy with
luminosity here is usually explained with the onset and growth of the 
accretion column at high luminosities \citep{BS76}. In this scenario, the height
of the column, and, therefore, the average displacement of the emission and the
line-forming regions from the   neutron star surface increase with luminosity. 
The magnetic field weakens rapidly with distance from the neutron
star and, therefore, the  CRSF should shift to lower energies. 
The problem is, however, that the predicted shift is much larger than the observed one. 
The column height depends on luminosity almost linearly \citep{BS76} and the magnetic field weakens with distance 
as $r^{-3}$, and yet brightening by more than an order of magnitude yields at most 25\% decrease
in the CRSF energy \citep{TsL2006,TsLS2010}. 
Moreover, large gradient of the magnetic field and of the accretion velocity are 
expected to smear out the line-like features making it difficult to explain why we observe CRSFs at all.

Several authors \citep{Burderi00,KW08}
considered variation of the magnetic field along the neutron star surface
as a possibility to explain the observed variation of the cyclotron line energy
with the pulse phase, which has similar magnitude as those associated with
the luminosity changes. Interaction of accreting plasma with the magnetosphere of
the neutron star defines the geometry of the emission region, so one could
imagine that a change in the accretion rate could also offset the location of the
polar cap, leading to a change in the observed CRSF energy. 
However, the observed variations would imply an unrealistically large shift of the 
hotspot location (by  50--60 degrees) from the magnetic pole. Furthermore, 
the observed change in the CRSF energy in this case would imply a dramatic change  
in the pulse profile shape, which is not observed.
  
 \begin{figure*}
\center{\includegraphics[width=1.0\linewidth]{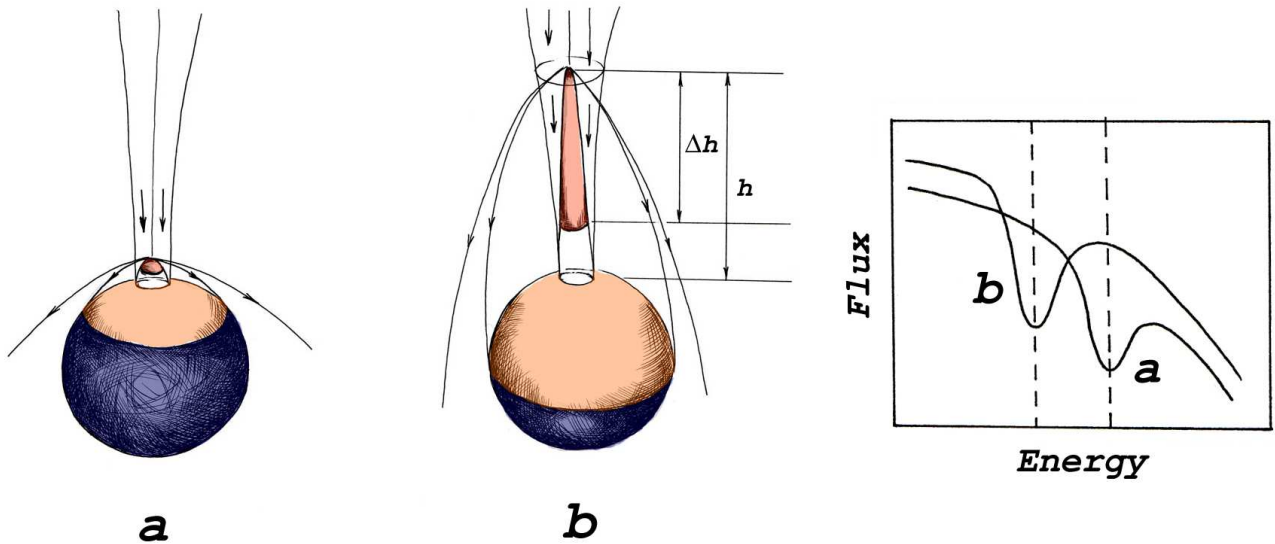}}
\caption{Sketch  of the accreting X-ray pulsar geometry, accretion column structure and the emergent spectrum.  
The larger is the accretion rate, the higher is the column, the larger illuminated fraction of the stellar surface, 
the weaker the average magnetic field, and the smaller the  cyclotron line energy.}
\label{fig:sketch}
\end{figure*}

On the other hand, a significant part of the column radiation should be intercepted by the stellar 
surface because of the relativistic beaming \citep{KFT76,LS88}. 
Small variations of the  $B$-field strength over the surface 
(only by a factor of two in the case of the dipole field) would imply that the 
variations of the typical cyclotron energy are small. 
Thus, it is more natural to assume that  the line is 
formed in the atmosphere of the neutron star due to reflection of the intercepted radiation. 
Increase in the mass accretion rate results in a larger luminosity, 
a higher column height and a larger illuminated part of the stellar surface (see Fig.~\ref{fig:sketch}). 
If the $B$-field decreases away from the magnetic poles, 
the cyclotron line energy should then decrease  with the luminosity and 
a negative correlation between the luminosity and the cyclotron line energy is reproduced. 
Here we discuss this scenario quantitatively 
and compare the model predictions with the data taking during 
a bright outburst of the transient X-ray pulsar V~0332+53.

\section{Model set up}
\label{sec:model}

Let us start from the physical picture of the accretion on the magnetized neutron star 
following earlier papers by \citet{BS76},  \citet{KFT76} and  \citet{LS88}.  
It is possible to distinguish the two regimes of accretion onto magnetized neutron stars 
depending on the mass accretion rates. 
At low accretion rate, free-falling protons heat part of the neutron star 
surface near its magnetic poles, and these bright spots radiate energy in the X-ray range. 
At high accretion rate, radiation pressure becomes significant and stops the infalling material
above the neutron star surface in the radiation-dominated shock. 
Below the shock, the matter slowly sinks down as the excess emission supporting the column escapes 
through the side walls. The column is expected to arise as soon as the
luminosity exceeds a critical value \citep{BS76}:
\begin {equation} \label{eq:limLum}
L^{*}\approx 4\times10^{36}\left(\frac{\kappa_{\rm T}}{\kappa_{\parallel}}\right)
\left(\frac{5l}{R}\right)
\left(\frac{M}{M_{\odot}}\right)\ \ergs,
\end{equation}
where $\kappa_{\parallel}$ is the electron scattering opacity along the magnetic field,
$\kappa_{\rm T}$ is the Thomson opacity, $M$ and $R$ are the mass and the radius of the star. 
The footprint of the accretion column at the stellar surface is normally a thin ring or an arc, and
$l$  is either circumference or the length of this arc. 
It is worth noticing that the optical depth across the column is of the order $L/L^*$ and  $L^{*}$ is 
much smaller than the Eddington luminosity for the whole star, but 
much larger than the Eddington luminosity scaled to the area of the  footprint of the accretion column.  
It depends on the accretion flow geometry and the magnetic field strength. 

The column height  depends on the  accretion rate $\dot{m}$ \citep{BS76,LS88}:
\begin {equation} \label{eq:ColHeigh}
\frac{h}{R}=\dot{m}\ \ln\left(\eta\frac{1+\dot{m}}{\dot{m}^{5/4}}\right),
\end{equation}
where
\begin{eqnarray} \label{eq:eta}
\eta&=&\left(\frac{B^2 d^2\kappa_{\parallel}}{7\pi c\sqrt{2GMR}}\right)^{1/4}  \nonumber \\
&=& 16 \left(\frac{B}{5\times 10^{12}\ \mbox{G}} \right)^{1/2} \left(\frac{d}{100\ \mbox{m}}\right)^{1/2} 
\left(\frac{\kappa_{\parallel}}{0.4} \right)^{1/4} ,
\end{eqnarray}
$d$ is the thickness of the accretion arc, and  
$\dot{m}=L/L^{**}$ is a ratio of the X-ray pulsar luminosity to the limiting 
luminosity for the magnetized neutron star  
\begin {equation} \label{eq:EddLum}
L^{**}\approx  10^{39}
\left(\frac{l/d}{50}\right)
\left(\frac{\kappa_{\rm T}}{\kappa_{\parallel}}\right)
\left(\frac{M}{M_{\odot}}\right)\ \ergs ,
\end{equation} 
which corresponds to the column height of $h\sim R$. 
The height, where matter stops, varies inside the accretion channel and depends on the distance from
its borders because the radiation energy density drops off sharply towards the edge of the column \citep{LS88}.
The height has its maximum value near the middle of the channel and decreases towards the borders.
Therefore, the radiation from the already stopped matter should pass through an outer layer of plasma
falling  at nearly the free-fall velocity $\beta=v/c=
\sqrt{r_{\rm S}/r}$ (here  $r_{\rm S}=2GM/c^2$ is the  Schwarzschild radius). 

These layers of the column are not supported by the radiation pressure, but 
are decelerated by radiative friction as discussed by \citet{LS88}. 
Our assumption of the free-fall velocity of the outer layers probably 
overestimates  their  actual mean velocity, but the general properties of the model are not affected.  
The optical thickness of  these layers is high enough to change significantly the angular distribution of the 
emergent radiation. 
As a result, the radiation is directed mainly towards the stellar surface due to the relativistic beaming.  
For the electron-scattering dominated column, the angular distribution of the column luminosity in the 
laboratory frame is given by (see Appendix \ref{app:ls88} and \citealt{KFT76,MT78}):
\begin {equation} \label{eq:LSd}
\frac{dL (\alpha) }{d\cos\alpha} = 
 I_0 \ \frac{D^4}{\gamma} 2 \sin\alpha  
\left(  1 +\frac{\pi}{2} D \sin\alpha\right),
\end{equation}
where $\alpha$ is the angle between the photon momentum and the velocity vector,  
$\gamma=1/\sqrt{1-\beta^2}$ is the Lorentz factor,  
$D=1/[\gamma(1-\beta\cos\alpha)]$ is the Doppler factor, and $I_0$ is the normalization constant. 
The emission pattern becomes more isotropic at large $r$, because of
the radial dependence of the free-fall velocity.

\begin{figure}
\center{\includegraphics[ width=0.95\linewidth]{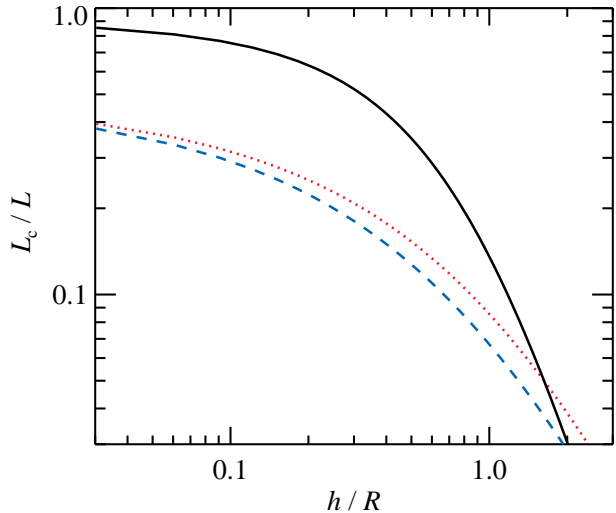}}
\caption{Fraction of the captured radiation from a point source above the neutron star surface
as a function of the height-to-radius ratio. 
The dashed blue curve corresponds to the isotropic source in flat space-time, 
the dotted red curve is for the isotropic source accounting for light bending in  Schwarzschild metric, 
and the solid black  curve is for the emission pattern given by equation  (\ref{eq:LSd}) 
and accounting for light bending.  Here $R=3 r_{\rm S}$.}
\label{fig:Part}
\end{figure}

The energy release and the distribution of the emerging radiation flux along the accretion column 
depends on the details of radiation--matter interaction and is a complicated radiation-hydrodynamical problem. 
Previous attempts  \citep[e.g.][]{BS76,BAK91,BW07} considered  one-dimensional models 
and/or did not account for the influence of the free-falling plasma. 
The actual problem is essentially 2D \citep{LS88} and has not been solved yet. 
For simplicity, we approximate the geometry of the accretion column by a thin stick on the magnetic pole. 
This approximation is reasonably accurate for high $B$-field pulsars up to $h\lesssim R$.
Here we also assume that most energy is emitted in a region of characteristic scale $\Delta h$ 
situated above the neutron star surface at height $h$ (see Fig.~\ref{fig:sketch}), so $\Delta h/h=1$  
for a homogeneously emitting column, or $\Delta h/h=0$ if all the energy is emitted 
within a thin shock at the top of the column.

Results weakly depend on the compactness of the star, which we fix here at $R/ r_{\rm S} =3$ 
(corresponding to  the radius of 13 km  for a $1.5\msun$ neutron star,
consistent with the recent measurements from X-ray bursters, 
see e.g. \citealt{SPRW11}).

\section{Reflected fraction and surface flux distribution}
\label{sec:refl}

At a high accretion rate, the accretion column radiates the released gravitational
energy of the falling matter through its sides. Part of the radiation leaves the system directly and part 
of it is captured by the stellar surface. The fraction of the captured radiation $L_{\rm c}/L$ 
depends on the height of the column, brightness distribution over the column, 
radiation beam pattern and the compactness of the star.
In  Schwarzschild metric,  for an isotropic, point-like source at radius $r$, 
it is given by a simple formula (see Appendix \ref{app:bend}): 
\begin{equation}
\frac{L_{\rm c}(r)}{L} = \frac{1}{2}\left( 1 - \cos\alpha_{\max}\right), 
\end{equation}
with 
\begin{equation}
\sin\alpha_{\max} = \frac{R}{r} \sqrt{\frac{1-r_{\rm S}/r}{1-r_{\rm S}/R}}.  
\end{equation}
In case of flat space-time, we can substitute $r_{\rm S}=0$. 
The captured fraction is slightly higher when light bending is accounted for. 
It drops from the maximal value of 0.5 to 0.1  when the source rises to $h\sim R$ (see Fig.~\ref{fig:Part}).

\begin{figure}
\center{\includegraphics[width=0.95\linewidth]{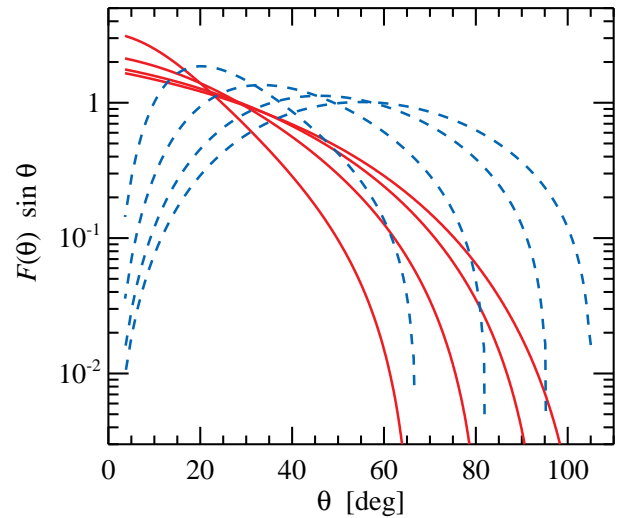}}
\caption{Dependence of the normalized flux  on co-latitude 
for an accretion column emitting according to the law (\ref{eq:LSd})  
with constant luminosity per unit height (i.e. constant $I_0$). 
The flux is normalized as $\int F(\theta)\sin\theta d\theta=1$.
The red solid lines correspond to $\Delta h/h=1$ and 
the blue dashed lines to $\Delta h/h=0$. 
Different lines show the distribution for various column heights: 
$h/R=0.5, 1, 2$ and $4$ (from left to right). Here $R=3 r_{\rm S}$. 
}
\label{fig:Hcol}
\end{figure}

For the emission pattern given by equation (\ref{eq:LSd}) and used  in the following calculations, 
the captured fraction increases even further, because of the strong beaming towards the star 
(see top curve in Fig.~\ref{fig:Part}).  
For a higher source position, $L_{\rm c}/L$ drops below the isotropic case at 
$h\sim 2R$, because of a rather small beaming resulting from a smaller free-fall velocity and 
a factor $\sin\alpha$, which reduces the radiation directed along the column.

Variations in the emission height lead also to a dramatic change in the distribution of the 
captured flux $F(\theta)$ over the stellar surface. 
We illustrate this effect here by considering a homogeneously emitting column (i.e. $I_0=\mbox{const}$ 
in equation~\ref{eq:LSd}) of various heights and $\Delta h/h$ (see Fig.~\ref{fig:Hcol}). 
The details of the calculations can be found in Appendix~\ref{app:bend}. 
For a low column, most of the emission hits the surface in the direct vicinity of the column 
at a co-latitude $\theta\lesssim\sqrt{2h/R}$ and the flux drops rapidly with distance from the column.  
For $\Delta h/h=1$, the flux $F(\theta)$ diverges as $1/\theta$, which is a consequence 
of  our assumption of an infinitely thin column. 
When $h\approx R$, the star is irradiated almost up to the equator and for even higher column, most 
of the stellar surface receives some flux. 
In a realistic pulsar, 
the captured flux is a sum of the contribution from the two antipodal columns.

\section{Formation of the cyclotron line}
\label{sec:line}
 
Radiation from the accretion column heats the  neutron star surface layers to the 
Compton temperature of radiation of a few keV. 
Under the  conditions of the neutron star atmosphere, 
the density is too small to produce any true absorption of the hard X-ray photons 
either in the continuum or at the cyclotron line energy. 
Therefore the impinging radiation is mostly reflected.\footnote{The absorption fraction 
as a function of energy is approximately (see e.g. appendix in \citealt{SMMH99}, or  sect. 2.3 of \citealt{Pou02})
$2\sqrt{k_{E}}/( \sqrt{k_{E}}+ \sqrt{k_{E} + \sigma_{\rm e}})$, 
where $\sigma_{\rm e}$ is the electron scattering opacity and  $k_{E}$ is a true absorption opacity 
(mainly free-free at  temperatures of a few keV).
A  pure hydrogen neutron star atmosphere with temperature $kT=1$ keV thermalizes 
about 4\% of an external blackbody flux with the color temperature of 5 keV. 
At $kT=3$ keV, the absorption fraction is about 1\% \citep{Doroshenko13}.}
The most important process affecting the spectrum of the reflected radiation is Compton scattering. 
The cross-section for Compton scattering in strong magnetic field is 
energy-dependent and has strong  resonances at  
\begin {equation} 
\frac{E^{(n)}_{\rm res}(B)}{m_{\rm e}c^2 } \! =\!  
\left\{  \begin{array}{ll}
\strut\displaystyle \! \! 
\frac{\sqrt{1+2nb\sin^2\xi}-1}{\sin^2\xi}, & \mbox{for}\ \xi\neq0, \  n=1,2,..., \\
\! \!  b ,                                             & \mbox{for}\ \xi=0, 
 \end{array} \right. 
\end{equation} 
where $b\equiv B/B_{\rm cr}$ is the $B$-field strength in units of the critical field strength 
$B_{\rm cr}= m_{\rm e}^2 c^3/e\hbar  = 4.412\times 10^{13}$ G, $\xi$ is the angle
between the field and photon momentum, and $m_{\rm e}$ is the electron mass. 
For $b\ll 1$, the resonance energies are $E^{(n)}_{\rm res} \approx n b  m_{\rm e}c^2$. 

Photons at the resonance energies cannot penetrate deep into the atmosphere, they interact in the surface 
layers and scatter back. 
In the cyclotron line wings, photons penetrate deeper into the atmosphere and scatter there with some
energy shift. If they scatter into the resonance energy, they cannot leave the atmosphere because of 
the larger optical depth there and escape instead in the line wings. 
Thus, the lack of the photons near the resonance is not filled in.
The absorption feature  at the resonance energy and the emission features 
in the wings appear  in the spectrum of the reflected radiation. 
The energy separation between the emission peaks can reach many Doppler widths (see 
\citealt{Avrett65}, \citealt{AH65}, Sect. 6.4 of \citealt{Ivanov73}  and \citealt{Mihalas1978}).   
Because the emission lines are broad, they merge with the  continuum 
and cannot be easily separated.  
For resonant Compton scattering in magnetic field, the line is not symmetric and 
the red wing of the emission line  is stronger, because of the recoil \citep{WS80,AH99,AH00,HL06}, 
which gives the relative photon energy shift of about $\Delta E/E\sim - E_{\rm res}/m_{\rm e}c^2\approx - b$.

The absorption features at the harmonic energies can be even stronger, because the absorbed 
photons are mostly reemitted at the energy of the fundamental. 
The typical X-ray pulsar spectrum cuts off at $\sim$30--50 keV and therefore 
the contribution of the harmonics will be negligible in high-field pulsars (as is the case of V~0332+53). 

Neglecting the asymmetry in the line shape, the  
centroid of the CRSF in the  reflected spectrum averaged over the surface and all angles is
determined by the $B$-field strength weighted with the
distribution of flux, $F(\theta)$ (Fig.~\ref{fig:Hcol}), 
and the  line equivalent width, $\mbox{EW}(\theta)$,  
over the neutron star surface:
\begin{equation} \label{eq:e_cycl}
%\langle E_{\rm cycl} \rangle=
 E_{\rm cycl} = \frac{m_{\rm e}c^2}{1+z}  \ 
\frac{\int\limits_{0}^{\pi}b(\theta)F(\theta)\ \mbox{EW}(\theta)\ \sin\theta\ d\theta}
{\int\limits_{0}^{\pi}F(\theta)\ \mbox{EW}(\theta)\ \sin\theta\ d \theta},
\end{equation} 
where $z$ is the gravitational redshift. 
Assuming that the line EW in the reflected radiation is constant over  the surface, we 
can relate the cyclotron line centroid to the mean field as 
\begin{equation} \label{eq:E_ave}
E_{\rm cycl} = \frac{m_{\rm e}c^2}{1+z} \frac{ \langle B\rangle}{B_{\rm cr}} , 
\end{equation} 
where
\begin{equation} \label{eq:b_ave}
\langle B\rangle=\frac{\int\limits_{0}^{\pi}B(\theta)F(\theta)\sin\theta\ d\theta}
{\int\limits_{0}^{\pi}F(\theta)\sin\theta\ d \theta}.
\end{equation} 
Variations of the magnetic field over the surface lead to the smearing of the line and 
its minimum width is then related to the magnetic field standard deviation: 
\begin{equation} \label{eq:sigmab}
\sigma_B^2  =\frac{\int\limits_{0}^{\pi}\left[ B(\theta)  - \langle B \rangle  \right]^2  F(\theta)\sin\theta\ d\theta}
{\int\limits_{0}^{\pi}F(\theta)\sin\theta\ d \theta}.
\end{equation} 
If  $(\sigma_B/\langle B \rangle)\times E_{\rm cycl}$ is smaller or comparable to the separation between 
the emission peaks, the CRSF will remain strong in the total spectrum of the reflected radiation.

\begin{figure}
\centering
\includegraphics[width=0.93\linewidth]{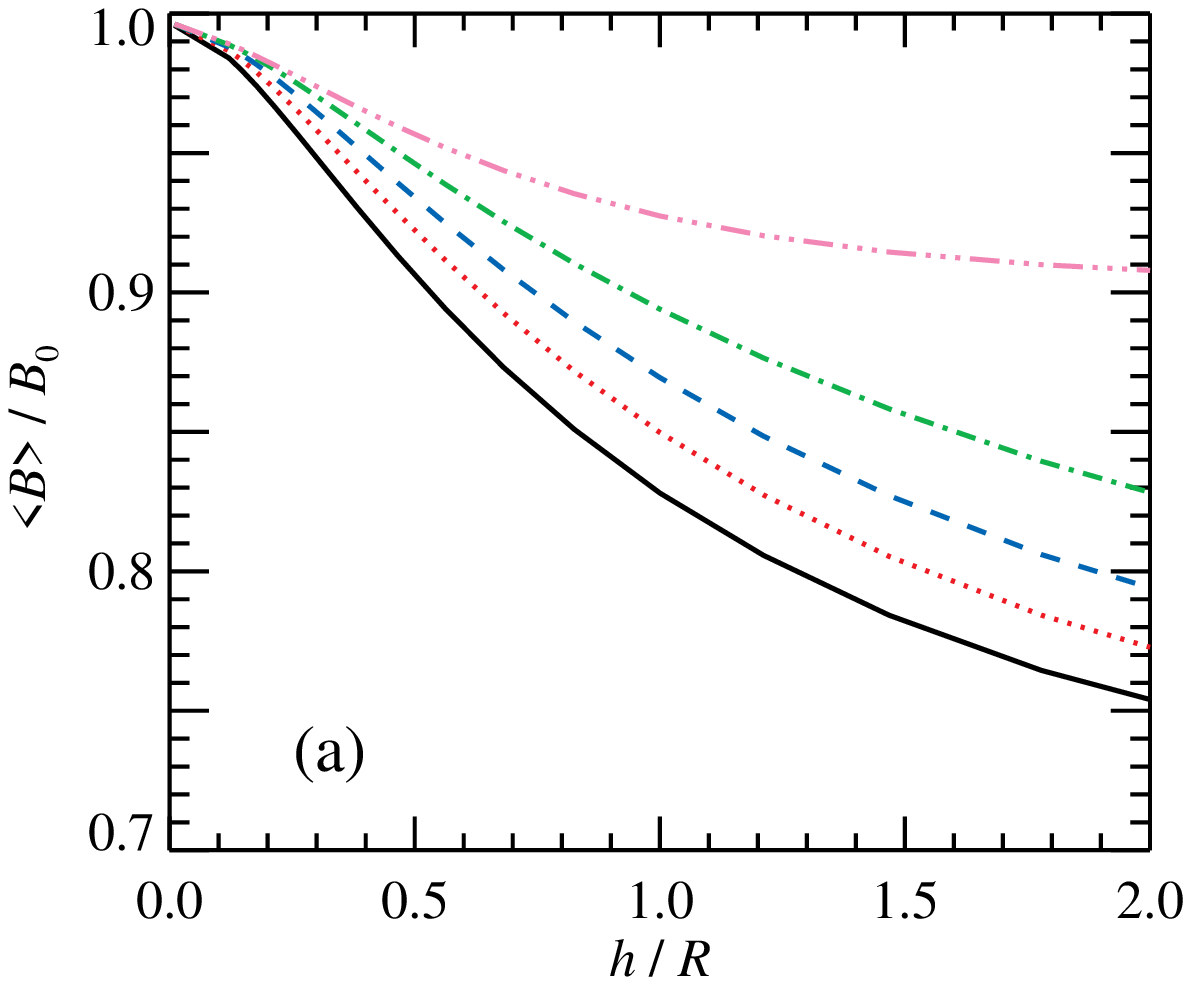}
\includegraphics[width=0.93\linewidth]{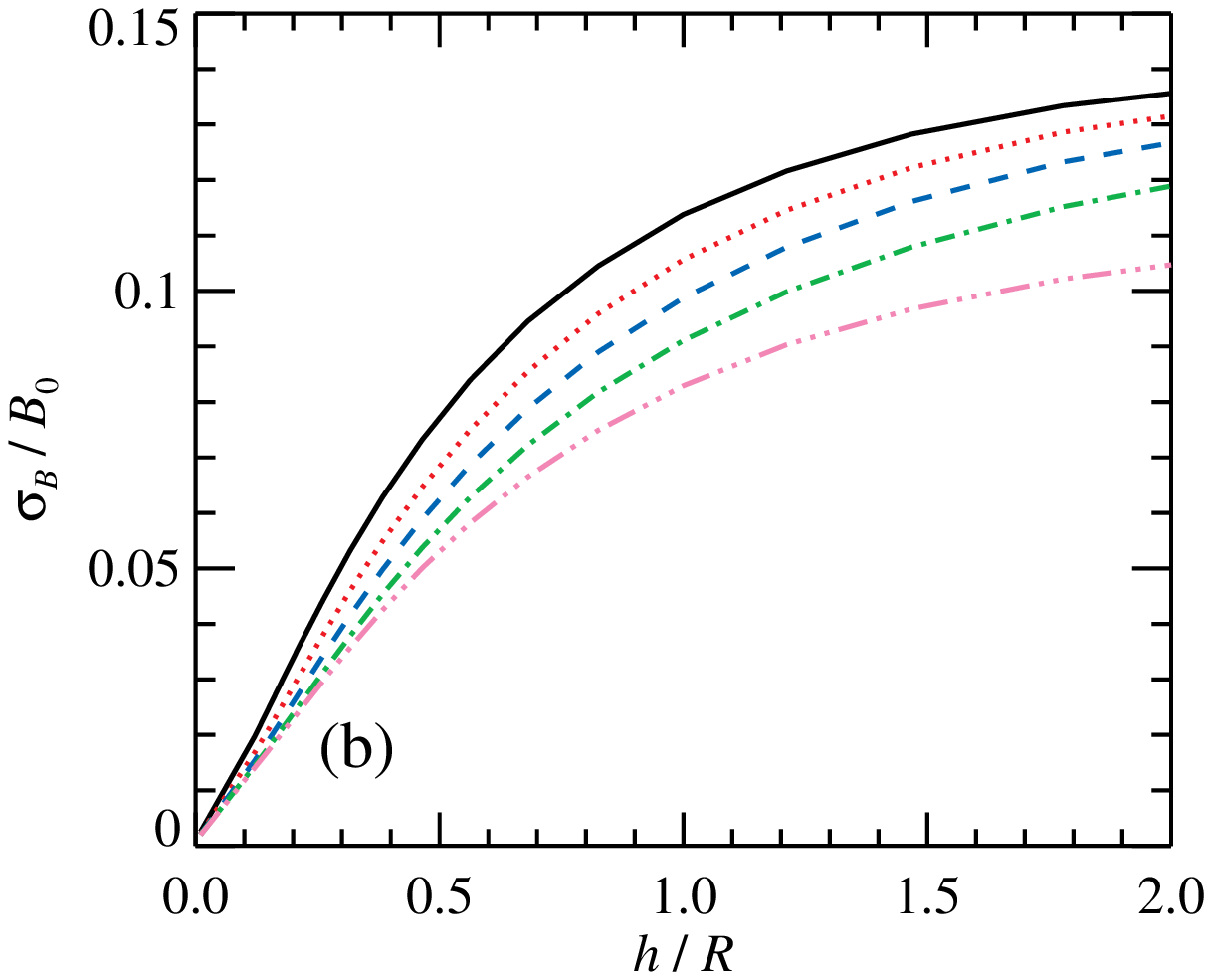}
\includegraphics[width=0.93\linewidth]{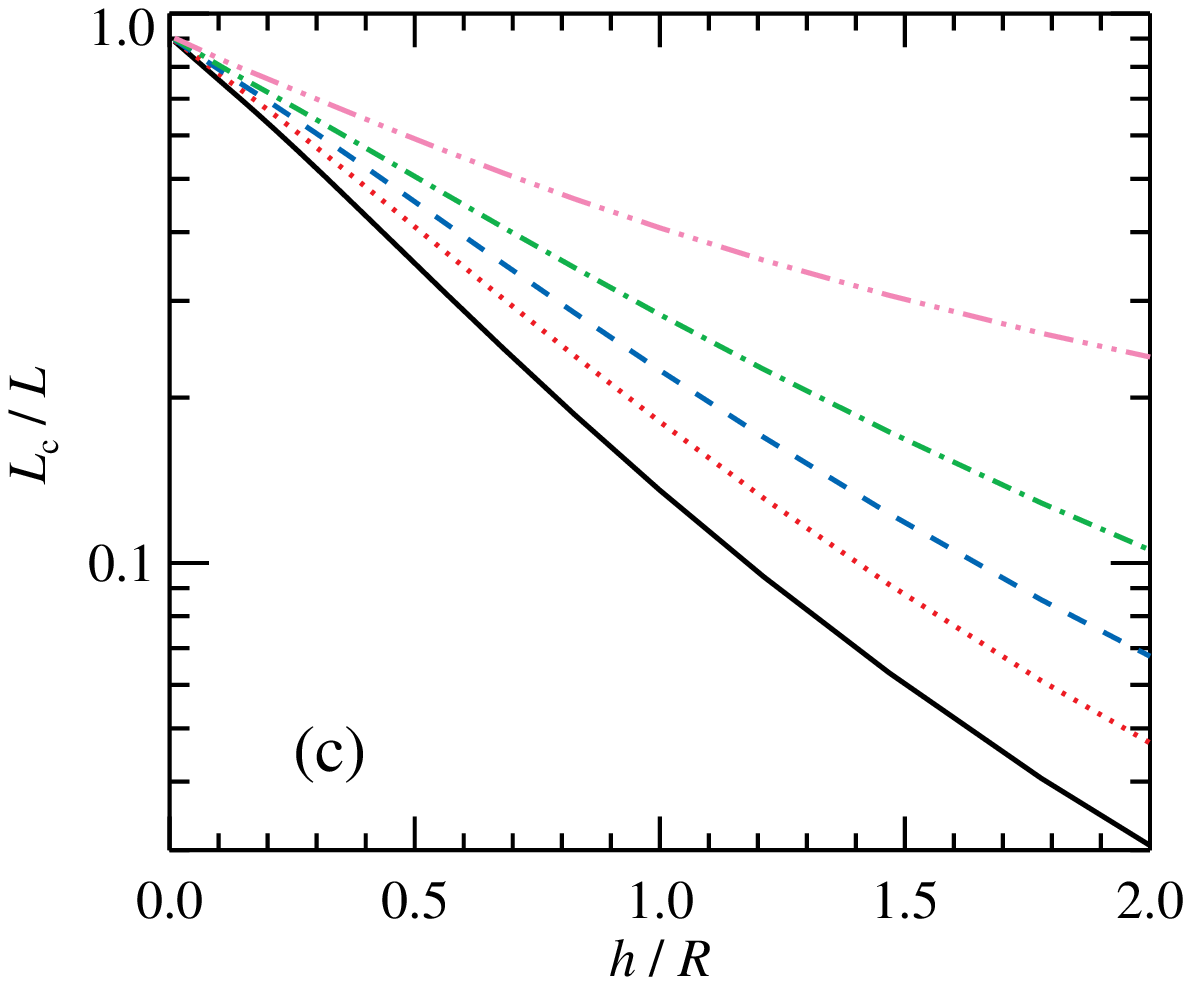}
 \caption{Theoretical dependences of (a) the mean magnetic field in the reflecting zone 
 for dipole field (in units of polar field) as defined by equation (\ref{eq:b_ave}), 
(b) the magnetic field standard deviation given by equation (\ref{eq:sigmab}), 
and (c) the fraction of the captured luminosity on the $h/R$ ratio for $R=3r_{\rm S}$. 
Black solid, red dotted, blue dashed, green dot-dashed and pink triple-dot-dashed line correspond to 
$\Delta h/h$=0, 0.3, 0.5, 0.7, 1.0, respectively. 
}
\label{fig:theor}
\end{figure}

At low accretion rate, the column is very low, $h/R\ll 1$, and only the polar region is illuminated, resulting 
in the line energy $E_{\rm cycl} =E_0$ corresponding to the polar field $B_0$.  
Because for the dipole field, the $B$-field strength drops only by a factor of two from the pole to the equator,
there is an obvious limitation for changes in the cyclotron line energy: the line
should lie in the energy interval $[E_0/2;E_0]$. 
A more realistic lower limit on the line energy can be obtained 
assuming a  uniformly illuminated surface, i.e. $F(\theta)={\rm const}$. 
In that case
\begin {equation}
\langle B\rangle_{\min}=\frac{1}{4}B_0\int\limits_{0}^{\pi}\sin\theta\sqrt{1+3\cos^2\theta}\ d\theta\simeq0.7 B_0.
\end{equation} 
Therefore, the line should not change its energy significantly and the model predicts the range for 
the line centroids of $[0.7E_0;E_0]$ in agreement with observations  \citep{TsL2006,TsLS2010}. 

In Fig.~\ref{fig:theor}, we present theoretical dependences of the mean magnetic field, standard deviation, and 
the captured fraction on the height of the column for different values of  $\Delta h/h$. 
These quantities will have a similar behavior as a function of luminosity too. 
We see that for the point-like source, variations in all quantities are the largest, 
while for a homogeneously emitting column, they are the smallest. 
If the whole column emits (i.e.  $\Delta h/h=1$), 
the polar regions are always strongly illuminated and $\langle B\rangle$ varies by only 10\%. 
On the other hand, for $\Delta h/h=0$, the mean field varies by at least 25\%.  
A negative correlation between the column height (luminosity) and the cyclotron line energy is produced. 
The smearing of the line, because of the magnetic field variation, does not reach more that 10--15\%  (see Fig.~\ref{fig:theor}b). 
Thus if the emission peaks in the reflected radiation are separated by more than 20\%, the CRSF would remain strong
even at high $h/R$. 
The strength of the CRSF in the total spectrum depends also linearly 
on the captured  luminosity fraction and 
is expected to decrease at high column  (see Fig.~\ref{fig:theor}c); therefore, a correlation 
between the line equivalent width and the energy is expected.

\section{Comparison with  observations}
\label{sec:obs}

We would like to compare our model with the data from the X-ray pulsar V~0332+53 obtained 
with the \textit{RXTE} and \textit{INTEGRAL} observatories during outburst in 2004--2005 
\citep{TsL2006,TsLS2010}. This source shows negative
correlation between the luminosity and the energy of the cyclotron line. It is the only data set which 
has the information about the behavior of the object in a so wide range of the luminosities,
from $\sim10^{37}$ up to $\sim4\times10^{38}\ \ergs$.

For a given parameter of the accretion column $\Delta h/h$, we compute 
the distribution of the reflected flux $F(\theta)$ as a function of the column height $h$. 
This is then converted to the average magnetic field $\langle B\rangle$ in units of the polar one $B_0$ 
using equation (\ref{eq:b_ave}) assuming a dipole field 
and to the cyclotron line energy $E_{\rm cycl}$ 
in units of the  polar value $E_0$ using equation (\ref{eq:E_ave}). 
Transition from the dependence on $h$ to the dependence on the luminosity is   
made as discussed in Sect.~\ref{sec:model} 
using equation (\ref{eq:ColHeigh}), where $\eta$ and $L^{**}$ are the parameters.

\begin{figure}
\centering
\includegraphics[width=0.95\linewidth]{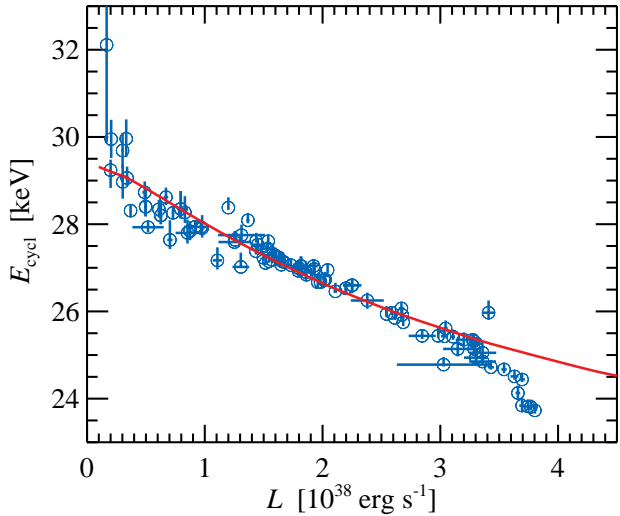}
\caption{Dependence of the cyclotron line energy on the luminosity in 
the X-ray pulsar V~0332+53 (circles with the error bars; from \citealt{TsLS2010}) and 
the best-fit theoretical relation (solid line) for $\Delta h/h=0.1$,  $\eta=15$, 
$E_0=29.5$~keV, and $L^{**}=21\times 10^{38}\ \ergs$. 
}
\label{fig:col_LS}
\end{figure}

\begin{deluxetable}{cccc}
\tabletypesize{\scriptsize}
\tablecaption{Best-fit parameters \label{tab:fits}}
%\tablewidth{\linewidth}
\tablehead{
\colhead{$\Delta h/h$ } & \colhead{$E_0$} & \colhead {$L^{**}$}  & \colhead{$\chi^{2}$/dof}  \\
    & (keV)  & $(10^{38}\ \ergs)$ &                                    } 
\startdata
 0.0  &    29.5$\pm$0.2  &    22.3$^{+2.4}_{-1.7}$   &    96.4/94 \\
 0.1  &    29.5$\pm$0.2  &    20.9$\pm$1.8  &    96.4/94 \\
 0.2  &    29.5$\pm$0.2  &    19.4$^{+1.9}_{-1.5}$   &    97.3/94 \\
 0.3  &    29.6$\pm$0.2  &    17.8$\pm$1.6  &    98.5/94 \\
 0.4  &    29.6$\pm$0.2  &    16.1$\pm$1.6  &   100.0/94 \\
 0.5  &    29.6$\pm$0.2  &    14.3$\pm$1.6  &   103.7/94 \\
 0.6  &    29.6$\pm$0.3  &    12.4$^{+1.4}_{-1.8}$ &   108.4/94\\
 0.7  &    29.7$^{+0.5}_{-0.3}$  &    10.2$^{+1.6}_{-2.0}$   &   116.7/94 \\
 0.8  &    29.9$^{+1.0}_{-0.3}$  &     7.6$^{+1.1}_{-3.2}$    &   130.6/94 \\
 0.9  &    32.8$^{+0.3}_{-0.7}$  &     1.1$^{+0.35}_{-0.13}$ &   124.2/94 \\
 1.0  &    28.7$^{+0.3}_{-0.3}$ &    12.5$^{+2.4}_{-2.8}$  &   227.3/94
\enddata
%\tablenotetext{a}{For 94 degrees of freedom. }
\end{deluxetable}

The cyclotron-line energy dependence on the luminosity for the X-ray pulsar V~0332+53 
\citep{TsLS2010} is shown in Fig.~\ref{fig:col_LS}. 
It is clear that the data at luminosities above $3.6\times 10^{38}\ \ergs$ cannot 
be described by our model and we neglect them in the fits. 
The remaining data are fitted with two free parameters $E_0$ and $L^{**}$. 
We fix $\eta=15$ (see equation \ref{eq:eta}), because the results depend on that parameter very weakly
and vary $\Delta h/h$ in the range between 0 and 1 with the step 0.1. 
Taking the errors on measured  $E_{\rm cycl}$ at their face values, 
gives the reduced $\chi^2$ significantly above unity, because of  
a  large spread of the line energies at a fixed luminosity. 
Therefore, we add a  systematic error of 0.15 keV in quadrature to the statistical errors of $E_{\rm cycl}$. 
A good agreement with the data, with the $\chi^2$ values below 100 (for 94 degrees of freedom),
is achieved for  $E_0\approx29.5$~keV, $\Delta h/h\lesssim 0.4$ and 
$L^{**}\approx 2\times 10^{39}\ \ergs$ (see Table~\ref{tab:fits}). 
The $\chi^2$ grows with  $\Delta h/h$ and $L^{**}$ decreases from about $2.2\times 10^{39}\ \ergs$
to $10^{39}\ \ergs$ when  $\Delta h/h$ changes from 0 to 0.7. 
The values of $L^{**}$ obtained in the fits should be taken as the upper limits on the actual value given 
by equation (\ref{eq:EddLum}), 
because the radiation from the upper part of the column might be completely blocked by the 
falling material and photons escape only at a height significantly smaller than the top of the 
accretion shock. 

The fitting procedure seems to indicate that the best description of the data is achieved for a column 
with most of the emission coming from its top. This does not necessarily mean that the lower part is 
not emitting, but rather that this emission does not hit the neutron star surface. 
It is actually expected, because in the lower part of the column 
the optical depth of the falling plasma above the shock 
becomes smaller  and the photons from the shock freely escape to the observer. 
Furthermore, our results are based on the assumption of the constant with latitude equivalent width. 
In reality, the scattering opacity is a strong function of the angle between 
the photon momentum and the magnetic field \citep[see e.g.][]{Pavlov80,SPavW12}, which 
necessarily will result in some latitude dependence and possibly will increase 
the line  EW of the spectrum reflected from the equatorial region. 
In that case, our result of the top-dominated emission from the column might 
be an artifact of this effect. 
 
The best-fit parameter indicate that the height of the column reaches $R$ at 
$L\sim 4 \times 10^{38}\ \ergs$. This is close to the luminosity, where large 
deviations of the model from the data are visible. 
The possible reason for this mismatch is that our approximation of the 
accretion column by a thin stick breaks down. 
In addition, at such large $L$,  radiation from the high anti-podal 
column starts to hit the equatorial region and becomes observable. This reduces the 
average ``observed'' $B$-field.  
In our model, the presented quantities are averaged over all observer angles and this 
effect is not accounted for.  

In our model, we assumed that the cyclotron line is formed exclusively by reflection. 
In reality, the lower parts of the accretion column can also contribute. 
Here the gradient of the magnetic field is not very large so that the line might not be smeared much 
and the radiation from the settling matter can penetrate through the free-falling 
gas, which is there already optically thin. Thus, we expect a complex interplay between 
the lines formed in the column and the surface, that can affect the resulting line centroid.  
Because no realistic 2D accretion column models have been published up to date,
the answer is highly uncertain and outside of the scope of the present paper.

%%%%%%%%%%%%%%%%%%%%%%%%%%%%%%%%%%%%%%%%%%%%%%%%%%%%%%%%%%%%%%%%%%%%%%%%%%%%%%
%% DISCUSSION AND/OR CONCLUSIONS                                            %%
%%%%%%%%%%%%%%%%%%%%%%%%%%%%%%%%%%%%%%%%%%%%%%%%%%%%%%%%%%%%%%%%%%%%%%%%%%%%%%
 
\section{Summary}
\label{sec:summary}

In this paper, we have proposed a  reflection model for the cyclotron line formation in X-ray pulsars. 
At high accretion rates, the accretion column is predicted to have a significant height 
illuminating a large fraction of the neutron star surface. 
Strong beaming of the column radiation towards the surface  
by the outer layers of free-falling plasma  guaranties that a significant 
fraction of the column emission is reflected from the surface.  

We argued that the reflected spectrum should have a strong CRSF at the energy 
depending on the local magnetic field strength.    
Small variations of the magnetic field along the surface imply that the line 
centroid energy can vary by at most 30\%, from the value corresponding to the field 
at the pole to the whole surface-averaged field.   
Changes in the pulsar luminosity are expected to correlate with the 
illuminated fraction of the stellar surface, and 
anti-correlate with the average magnetic field and, therefore, with the line centroid energy, 
exactly as observed during the outburst of the X-ray pulsar V~0332+53. 
Our model has profound  implications for the interpretation of the data on the 
cyclotron lines observed in X-ray pulsars. 

In order to predict the line parameters more accurately, 
a detailed model of the reflection of the column radiation from the atmosphere is required. 
It necessarily should include the dependence on the orientation of the local $B$-field 
as well as the angles of the incoming and reflected radiation. 
We would then be in a position to predict variations of the line energy and EW with the 
pulsar phase. This project is left for a future study.

\acknowledgments
This research was supported by 
the Magnus Ehrnrooth foundation (AAM), 
the Jenny and Antti Wihuri foundation,
the Academy of  Finland (grants 259490, 259284 and 270006), 
the German Research Foundation (DFG) grant SFB/Transregio 7 ``Gravitational Wave Astronomy''
and the Russian Foundation for Basic Research (grants 12-02-97006-r-povolzhe-a, VFS; 
11-02-01328, 12-02-01265, and 13-02-12094-ofim, AAL). 
We thank Alexander Serber for useful comments.

\appendix

\section{Emission from the rapidly falling plasma in an accretion column}
\label{app:ls88}

Let us consider an accretion column with the gas falling with velocity $\beta=v/c$. 
Let us assume that the angular distribution of radiation escaping from the surface 
in the gas comoving frame is given by the bolometric intensity 
\begin {equation}  
I'(\zeta')=I_0\ (1+a \cos\zeta'), 
\end{equation} 
where $\zeta'$ is the angle measured from the local normal to the 
surface, $I_0$ is a constant, and $a$ is the anisotropy parameter. 
In the calculations, we have chosen $a=2$, which corresponds to the case of electron-scattering dominated 
optically thick atmosphere \citep{Cha60,Sob63}, but keep here the formulae as general as possible. 
We define the coordinate system with the $z$-axis along the direction of motion of the gas,
and the $x$-axis along the normal to the surface. 
The four-vector of the photon momentum in the lab frame is 
$\underline{k} = k \left\{ 1, \bmath{\omega} \right\}$,
where $k=h\nu/m_{\rm e}c^2$ and $\bmath{\omega}$ is the unit vector along the photon momentum, which 
makes angle $\alpha$ with the  $z$-axis: 
$\bmath{\omega} = (\sin\alpha\cos\phi, \sin\alpha\sin\phi, \cos\alpha)$. 
Relative to the surface normal $\bmath{n}=(1,0,0)$, it makes angle $\zeta$, 
so that  $\cos\zeta=\bmath{\omega} \cdot \bmath{n} =\sin\alpha\cos\phi$.

The photon four-momentum in the frame comoving with the spot, 
$\underline{k'} = k' \left\{ 1, \bmath{\omega'} \right\}$, 
is obtained from the Lorentz transformation.  
The energy is transformed as $k'=k/D$ and the unit vector along photon momentum is 
\begin {equation}  
\bmath{\omega'}=\left(D\sin\alpha\cos\phi, D\sin\alpha\sin\phi, 
\frac{\cos\alpha-\beta}{1-\beta\cos\alpha} \right). 
\end{equation} 
Here the Doppler factor  is
\begin {equation}   \label{eq:dop}
D=\frac{1}{\gamma(1-\beta\cos\alpha)} ,
\end{equation} 
and $\gamma=1/\sqrt{1-\beta^2}$ is the Lorentz factor. 
Note that because of the relativistic aberration, the cosine of the 
projection angle we see the surface element that moves along its surface 
is  (see also \citealt{PG03,PB06}): 
\begin {equation}   \label{eq:coszeta}
\cos\zeta'=D \cos\zeta=D \sin\alpha\cos\phi . 
\end{equation} 
 
The bolometric power emitted by a unit area  surface element  in the comoving frame 
per unit solid angle is given by 
 \begin {equation}   \label{eq:power_com}
\frac{dL'}{d\bmath{\omega}'} =  \   I'(\zeta') \ \cos\zeta'. 
\end{equation} 
The angle-integrated emitted power is 
 \begin {equation}   \label{eq:tot_power_com}
L' = \oint \frac{dL'}{d\bmath{\omega} '} d\bmath{\omega} ' = \int_0^{2\pi} d\varphi' \ \int_0^1  \ I' (\zeta' )\ \cos\zeta'  \ d\cos\zeta'  
=I_0\ \pi \left(1+\frac{2}{3}a\right). 
\end{equation} 
In order to evaluate the flux on the neutron star surface, we need to know 
the emitted power in the lab frame, which is given by \citep{RL79} 
\begin {equation}   \label{eq:power_lab}
\frac{dL}{d\bmath{\omega} } = \frac{D^3}{\gamma}  \frac{dL'}{d\bmath{\omega}' }
= \frac{D^4}{\gamma}   I'(\zeta')\ \sin\alpha\cos\phi  . 
\end{equation} 
Note that here the first transformation does not involve factor $D^4$, because 
we consider a steady-state source, not a moving blob \citep{RL79,SMMP97}. 
For a thin accretion column, photons emitted at any azimuthal angles $\phi$ will have basically the same 
trajectory that is described just by angle $\alpha$. Thus, we can integrate over $\phi$ 
to obtain the final expression for the emission pattern 
 \begin {equation}   \label{eq:power_lab_alpha}
\frac{dL}{d\cos\alpha}= \int_{-\pi/2}^{\pi/2} d\phi \ \frac{dL}{d\omega}  = 
 I_0 \ \frac{D^4}{\gamma}\ 2 \ \sin\alpha \left( 1 + {a} \ \frac{\pi}{4} D \sin\alpha\right) . 
\end{equation} 
This expression is  different somewhat from that given by \citet{KFT76} and \citet{MT78} as it contains one less Doppler factor. 
For the emission forward-back symmetric in the comoving frame, 
the total emitted power is Lorentz invariant \citep{RL79}, which can be checked by integrating 
expression (\ref{eq:power_lab_alpha}): 
 \begin {equation}   \label{eq:tot_power_lab}
L = \int_{-1}^1  \frac{dL}{d\cos\alpha} d\cos\alpha =  L' . 
\end{equation}

\section{Photon propagation in the vicinity of the neutron star}
\label{app:bend}

\begin{figure}
\centerline{\epsfig{file=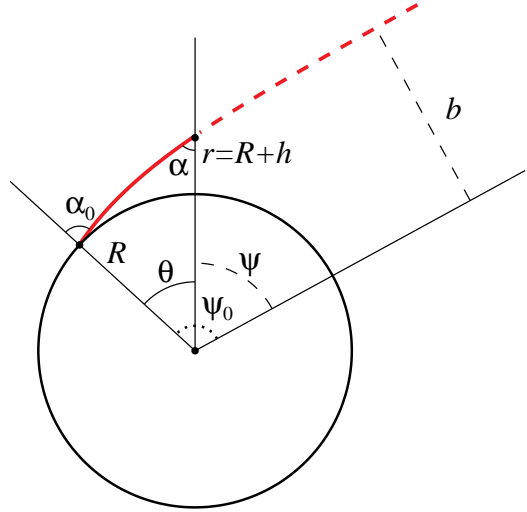,width=10.0cm}}
\caption{Geometry of light bending in Schwarzschild metric.  
The red solid line shows the photon trajectory from the emission point 
at distance $r=R+h$ to the stellar surface at radius $R$ and co-latitude $\theta$. 
The red dashed line is the continuation of the trajectory in the opposite direction 
to the infinity. 
\label{fig:bend_geom}}
\end{figure}

Figure \ref{fig:bend_geom} depicts the trajectory of a photon (red solid
line) which is emitted at the accretion column at height $h$
above the neutron star surface in the direction that makes angle 
$\alpha$ with the radial direction towards the stellar center. 
The photon's path, described by the distance $r=R+h$ and azimuthal angle $\psi$, obeys 
the equation of motion \citep[e.g.][]{MTW73,PFC83}: 
\begin{equation}\label{eq:motion}
\left( \frac{du}{d\psi} \right)^2 +  
(1-u) u^2 = \frac{1}{b^2}, 
\end{equation}
where $u=r_{\rm S}/r$ and $b$ is the impact parameter in units of  
the Schwarzschild radius $r_{\rm S}=2GM/c^2$ for a neutron star with mass $M$. 
The impact parameter and the angle, $\alpha$, between the radial direction
and the photon trajectory are related by \citep[e.g.][]{PFC83,B02}
\begin{equation}\label{eq:impact}
  b=\frac{\sin\alpha }{u\sqrt{1-u}} .
  %=   (1+z)  \frac{\sin\alpha}{u }  .
\end{equation}
The azimuthal angle $\psi$ measured 
from the continuation of the trajectory towards infinity (dashed red line in Fig.~\ref{fig:bend_geom}) 
can be  obtained from equation (\ref{eq:motion}):  
\begin{equation}\label{eq:bend1}
  \psi (r,\alpha)=\int_0^{u}  {d u'} \left[ b^{-2} - (1-u') u'^2
  \right]^{-1/2}.
\end{equation}
For $r$ larger than $2r_{\rm S}$ the  elliptic integral (\ref{eq:bend1}) 
can be approximated with a high accuracy by \citep{B02,PB06}  
\begin{equation}\label{eq:ABappr}
\cos\alpha = u(r) + [1-u(r)] \cos\psi . 
\end{equation}

%%%%%%%%%%%%%%%%%%%%%%%%%%%%%%%%%%%%%%%%%%%%%
%%%%%%%%%%%%%%%%%%%%%%%%%%%%%%%%%%%%%%%%%%%%%

The photon trajectory with impact parameter $b$ 
meets the neutron star surface of radius $R$ at an angle $\alpha_0$ 
relative to the normal, which is (see equation (\ref{eq:impact})):
\begin{equation}\label{eq:alpha0}
\sin\alpha_0 = \sin\alpha\ \frac{u(R) \sqrt{1-u(R)}}{u(r) \sqrt{1-u(r)}} . 
\end{equation}
The intersection point lies at the azimuth $\psi_0$ that can be obtained from 
equation (\ref{eq:ABappr}) substituting there compactness $u=u(R)=r_{\rm S}/R$ 
and $\alpha_0$ instead of $\alpha$.
 Equation (\ref{eq:alpha0}) immediately constrains the maximum angle $\alpha$, when the trajectory is 
still intercepted by the star,  corresponding to $\alpha_0=\pi/2$: 
 \begin{equation}\label{eq:alphamax}
\sin\alpha_{\max} = \frac{R}{r}\ \sqrt{ \frac {1-u(r)} {1-u(R)}} . 
\end{equation} 
For a source at distance $r$ from the star 
emitting according to the angular pattern $dL(\alpha)/d\cos\alpha$ with total luminosity $L$, 
we then can obtain the fraction of radiation captured by the star: 
\begin{equation}\label{eq:captured}
\frac{L_{\rm c}}{L} = \frac{\int_{\cos\alpha_{\max}}^1 \frac{dL(\alpha)}{d\cos\alpha} \ d\cos\alpha }
{ \int_{-1}^1 \frac{dL(\alpha)}{d\cos\alpha} \ d\cos\alpha } .
\end{equation} 
For an isotropic source, equation (\ref{eq:captured}) is reduced to 
\begin{equation}\label{eq:captured_iso}
\frac{L_{\rm c}}{L} = \frac{1}{2} \left( 1- \cos\alpha_{\max} \right) .
\end{equation} 
   
Distribution of the intercepted luminosity over the neutron star surface is given by
\begin{equation}\label{eq:Ltheta}
\frac{dL_{\rm c}(\theta)}{d\cos\theta} = \frac{dL(\alpha)}{d\cos\alpha} \ 
\frac{d\cos\alpha}{d\cos\theta} \   \frac{dL_{\rm c}}{dL} ,
\end{equation} 
where the first factor on the rhs is the specified emission pattern.  
We compute the second factor numerically differentiating   $\theta=\psi_0-\psi$ as a function of $\alpha$, 
with both $\psi$ and $\psi_0$ obtained via approximate light bending equation (\ref{eq:ABappr}) and 
$\alpha_0$  from equation (\ref{eq:alpha0}). 
The last factor, $dL_{\rm c}/dL = [1-u(r)]/[1-u(R)]$,  
just contains two redshift factors  of the type $1+z=1/\sqrt{1-u}$: 
one from the photon energy change  
in the gravitational field and another one from the time dilation.

%%%%%%%%%%%%%%%%%%%%%%%%%%%%%%%%%%%%%%%%%%%%%%%%%%%%%%%%%%%%%%%%%%%%%%%%%%%%%%
%% Bibliography                                                             %%
%%%%%%%%%%%%%%%%%%%%%%%%%%%%%%%%%%%%%%%%%%%%%%%%%%%%%%%%%%%%%%%%%%%%%%%%%%%%%%
%\bibliographystyle{apj}
%\bibliography{allbib}

\end{document}